\documentclass[epj,1p]{svjour}
\usepackage{graphicx}% Include figure files
\usepackage{dcolumn}% Align table columns on decimal point
\usepackage{bm}% bold math
\usepackage{hyperref}% add hypertext capabilities
\usepackage{amssymb}%\usepackage[mathlines]{lineno}% Enable numbering of text and display math
\usepackage{hyperref}% add hypertext capabilities
\RequirePackage[normalem]{ulem} %DIF PREAMBLE
\RequirePackage{color}\definecolor{RED}{rgb}{1,0,0}\definecolor{BLUE}{rgb}{0,0,1} %DIF PREAMBLE
\begin{document}
%\date{}
%\unitlength=1.00mm \special{em:linewidth 0.4pt}
%\linethickness{0.4pt}
%\thispagestyle{empty}
\newcommand{\be}{\begin{equation}}
\newcommand{\ee}{\end{equation}}
\newcommand{\ba}{\begin{eqnarray}}
\newcommand{\ea}{\end{eqnarray}}
\newcommand{\Gam}{\Gamma[\varphi]}
\newcommand{\Gamm}{\Gamma[\varphi,\Theta]}
\thispagestyle{empty}

\title{Three strongly correlated charged bosons   in a
one-dimensional harmonic trap: natural orbital occupancies}

\author{ Przemys\l aw Ko\'scik,
Institute of Physics,  Jan Kochanowski University\\
ul. \'Swi\c{e}tokrzyska 15, 25-406 Kielce, Poland}
%\ead{koscik@pu.kielce.pl}

\abstract{We study a one-dimensional system composed of three
charged bosons
   confined in an external harmonic
 potential.  More precisely, we  investigate   the ground-state correlation properties of the
 system, paying particular attention to the
strong-interaction limit. We explain for the first time the nature
of the degeneracies appearing in this limit  in the  spectrum of the
 reduced density matrix. An explicit representation
of the asymptotic natural orbitals and their occupancies is given in
terms of some integral equations.}

\maketitle
\section{Introduction}
In recent years there has been a growing interest in systems of
interacting particles trapped in potential wells because of their
possible use in quantum information technology \cite{inf}.
Especially, systems composed of particles  held together in
 harmonic potentials have drawn considerable
theoretical attention
\cite{21,puente,sim,22,manz,lin3,monk,posi,posi1,kosPLA,kosl,kose,kose2,bos,bos1,del}\\\cite{del1,tg0,sun1,gi,bao,blu,nboson,gonz,ji,astr,cios}.
Besides fermionic systems, which serve well as  models of quantum
dots, many attempts have been made to explore the properties of the
bosonic ones with a contact potential
\cite{bos,bos1,del,del1,tg0,sun1,gi}. Such systems in the
one-dimensional (1D) limit have attracted   much attention and their
properties appear  to be   well understood
\cite{del,del1,tg0,sun1,gi}. Also, there has been considerable
 interest in  the properties of artificial atoms composed of
bosons interacting via a Coulomb potential
\cite{bao,blu,nboson,gonz,cios,ji,astr}, which  in turn serve well
as models of  electromagnetically trapped ions \cite{ion}. However,
there has been   so far relatively little theoretical research on
such systems  in the strictly  1D limit.

 In  the present paper we consider  an ideal 1D system
composed of three identical bosons described by the Hamiltonian \be
H=\sum_{i=1}^3-{1\over 2}{\partial^2\over
\partial x_{i}^2}+V,\label{hamk}\ee with %\begin{eqnarray} V={1\over 2}\sum_{i=1}^{3} x_{i}^2
 %+\sum_{i>j=1}^{3} {g\over |x_{i}-x_{j}|}=\nonumber\\={3\over 2} X^2+\sum_{i>j=1}^{3}[{1\over 6}
%x_{ij}^2+{g\over |x_{ij}|}]\label{pot},\end{eqnarray}
\begin{eqnarray} V={1\over 2}\sum_{i=1}^{3} x_{i}^2
 +\sum_{i>j=1}^{3} {g\over |x_{i}-x_{j}|},\label{ghjj}\end{eqnarray}
or equivalently  \be V={3\over 2} X^2+\sum_{i>j=1}^{3}[{1\over 6}
x_{ij}^2+{g\over |x_{ij}|}]\label{pot},\ee where
$x_{ij}=x_{i}-x_{j}$ are the interparticle distances and $X={1\over
3}(x_{1}+x_{2}+x_{3})$ is the centre-of-mass. In (\ref{hamk}), the
spatial variable $x$ is given in oscillatory units
$\sqrt{\hbar/m\omega}$, and $g$ is the ratio of the Coulomb and the
confinement energies: $g={e^2 \over \epsilon}\sqrt{{m\over
\hbar^3\omega}}$.

Experimentally, a 1D configuration of trapped ions can be realized
in a 3D harmonic trap with a transverse trapping frequency
$\omega_{\perp}$ much larger than the axial one $\omega$,
$\epsilon=\omega_{\perp}/\omega\gg 1$. Although the 1D Hamiltonian
(\ref{hamk}) is strictly valid only in the limit
$\epsilon\rightarrow\infty$, it works well even at moderate values
of $\epsilon$ if the confinement in the $x$ direction is very weak
($g\rightarrow\infty$). For more details on this point we refer the
reader to \cite{kose2}.

 The main goal of this paper is to  make a detailed investigation   of the ground-state correlation
properties of the  system  (\ref{hamk})  in the strong-interaction
limit ($g\rightarrow\infty$). To this end, we use an approximation
based on the second-order Taylor series expansion of (\ref{pot})
around the classical equilibrium
 distances of the particles \cite{cios}.
Within the framework of this approximation, an asymptotically exact
expression for the ground-state bosonic wavefunction at the
$g\rightarrow\infty$ limit
 can be obtained. We derive an explicit  expression for the asymptotic reduced density
matrix (RDM) and investigate for the first time the nature of the
degeneracies appearing in its spectrum. We provide an explicit
representation of the asymptotic natural orbitals and their
occupancies, given by integral equations independent of $g$. In
particular, we  show that only the three natural orbitals contribute
significantly to the asymptotic bosonic ground-state. Moreover,  to
gain insight into the general features  of the Schmidt expansion of
the RDM, we determine numerically the values of the three lowest
occupancies   over a wide range of values of $g$.

 This paper is  organized as follows. Section \ref{1}
derives     closed-form analytical approximate solutions for the
bosonic ground-state of the system
   (\ref{hamk}). Section \ref{test} tests their
  validity and provides, in particular,  detailed results
 for  the dependencies of the degree of correlation
    on  $g$.  Section
\ref{2} derives $g$-independent integral equations defining  the
asymptotic natural orbitals and their occupancies. Here, two
different
 forms of the  Schmidt expansion of the asymptotic RDM  will be discussed. Finally, some concluding remarks are placed in
Section \ref{3}.

\section{Harmonic approximation}\label{1}

 The potential (\ref{ghjj}) attains its minimum at six
points, namely $(x_{i}^{m },x_{j}^{m},x_{k}^{m})$,
$x_{i}^{m}=-x_{c}$, $x_{j}^{m}=0$, $x_{k}^{m}=x_{c}$, where
$x_{c}={(10 g)^{1\over 3}/ 2}$, and $\{i, j, k\}$ are the
permutations of $\{1, 2, 3\}$. Consequently, the
 distances between the three particles in the ground-state
of the system in the classical limit are $ x_{21}^{m}=
x_{2}^{m}-x_{1}^{m}=x_{c}={(10 g)^{1\over 3}/ 2},$ $
x_{32}^{m}=x_{3}^{m}-x_{2}^{m}=x_{c}= {(10 g)^{1\over 3}/ 2},$ $
x_{31}^{m}=x_{3}^{m}-x_{1}^{m}=2x_{c}={(10 g)^{1\over 3}}, $ and
$X^{m}=(x_{1}^{m}+x_{2}^{m}+x_{3}^{m})/3=0$, where we have referred
to the minimum $\{i=1,j=2,k=3\}$. The Taylor expansion of
(\ref{pot}) around $(x_{21}^{m},x_{32}^{m},x_{31}^{m},X^{m})$ is

\begin{eqnarray} V=V_{m}+{3\over 2}X^2+\nonumber\\+\sum_{i>j=1}^{3}{\partial V\over
\partial
x_{ij}}\mid_{x_{ij}=x_{ij}^{m}}(x_{ij}-x_{ij}^{m})+\nonumber\\+
\sum_{i>j=1}^{3}{1\over 2}{\partial^2 V\over
\partial x_{ij}^2}\mid_{x_{ij}=x_{ij}^{m}}(x_{ij}-x_{ij}^{m})^2+...+\end{eqnarray}
 Noticing that \be
\sum_{i>j=1}^{3}{\partial V\over
\partial
x_{ij}}\mid_{x_{ij}=x_{ij}^{m}}(x_{i}-x_{j}-x_{ij}^{m})=0,\ee
$({\partial V\over
\partial
x_{ij}}\mid_{x_{ij}=x_{ij}^{m}}=(-1)^{i+j}{7g^{1\over 3}\over 3
*10^{2\over 3}})$ and retaining the terms up to
second order, one gets the approximation
\begin{eqnarray}V\approx V_{ap}=V_{m}+{3\over 2}({{x}_{1}+{x}_{2}+{x}_{3}\over 3})^2+\nonumber\\+{29\over
30}(x_{2}-x_{1}- x_{c})^2+{29\over
30}(x_{3}-x_{2}-x_{c})^2+\nonumber\\+{4\over
15}(x_{3}-x_{1}-2x_{c})^2\label{roz},\end{eqnarray}wherein \be
V_{m}={3*5^{2\over 3} g^{2\over 3}\over 2^{4\over 3}}.\ee  Thus the
Hamiltonian (\ref{hamk}) is approximated by \be H_{ap}=
\sum_{i=1}^3-{1\over 2}{\partial^2\over
\partial x_{i}^2}+V_{ap}.\label{f1ff}\ee
It is convenient to introduce new coordinates in (\ref{f1ff}), \be
x_{1}\mapsto \tilde{x}_{1}- x_{c},x_{2}\mapsto\tilde{x}_{2}
,x_{3}\mapsto \tilde{x}_{3}+ x_{c},\label{coo}\ee  so that the
corresponding Schr\"{o}dinger equation takes the form
\begin{eqnarray} \{\sum_{i=1}^3-{1\over 2}{\partial^2\over
\partial \tilde{x}_{i}^2}+V_{m}+{3\over 2}({\tilde{x}_{1}+\tilde{x}_{2}+\tilde{x}_{3}\over 3})^2+\nonumber\\+{29\over
30}(\tilde{x}_{2}-\tilde{x}_{1})^2+{29\over
30}(\tilde{x}_{3}-\tilde{x}_{2})^2+\nonumber\\+{4\over
15}(\tilde{x}_{3}-\tilde{x}_{1})^2\}\psi_{ap}=E_{ap}\psi_{ap}.\label{fff}\end{eqnarray}
 Eq.
(\ref{fff}) would suggest an ansatz of the form
\begin{eqnarray} \psi_{ap}(\tilde{x}_{1},\tilde{x}_{2},\tilde{x}_{3})=\nonumber\\=e^{-\alpha ({\tilde{x}_{1}+\tilde{x}_{2}+\tilde{x}_{3}\over 3})^2}
e^{-\beta (\tilde{x}_{2}-\tilde{x}_{1})^2}e^{-\beta
(\tilde{x}_{3}-\tilde{x}_{2})^2} e^{-\gamma
(\tilde{x}_{3}-\tilde{x}_{1})^2} \label{roz1}.\end{eqnarray} After
substituting (\ref{roz1}) into (\ref{fff}), and performing some
straightforward  algebra,   we find parameter values
$\alpha,\beta,\gamma$ at which the function (\ref{roz1})
 satisfies Eq. (\ref{fff}), that is
$$\alpha={3\over 2},\beta={1\over 6} \sqrt{{29\over
5}},\gamma={1\over 60}(15\sqrt{3}-\sqrt{145}),$$ and \be
E_{ap}={1\over 20}(10+10\sqrt{3}+2\sqrt{145}+15*10^{2\over 3}
g^{2\over 3})\label{app}.\ee  Changing the variables back in
(\ref{roz1}) in accordance with (\ref{coo}), we can construct then
the approximate spatial symmetric
  wavefunction:
\begin{eqnarray}\psi_{B}^{ap}(x_{1},x_{2},x_{3})=C \sum_{\{i, j, k\}}\psi_{ap}(x_{i}+x_{c},x_{j},x_{k}-x_{c})
\label{pps},\end{eqnarray} where $\{i, j, k\}$ are again the
integers $\{1, 2, 3\}$ permuted into a different order. The
normalization constant in (\ref{pps}) can be calculated
analytically. However, since it is a quite lengthy formula,
 we report here only
 its value as $g\rightarrow\infty$  \be
C_{\infty}={({29\over 5})^{1\over 8}\over \sqrt{2}3^{3\over
8}\pi^{3\over 4}}.\ee

 Before going further it should be stressed that the
approximation
 (\ref{roz}) coincides with that obtained  from the second-order
Taylor series expansion of (\ref{ghjj}) around the classical
equilibrium positions of the particles $(-x_{c},0,x_{c})$. The
approximation strategy considered here yields thus the results
consistent with those of the standard normal-mode theory \cite{ji}.

\section{Numerical tests}\label{test}

\begin{figure}[h]
\begin{center}
\includegraphics[width=0.45\textwidth]{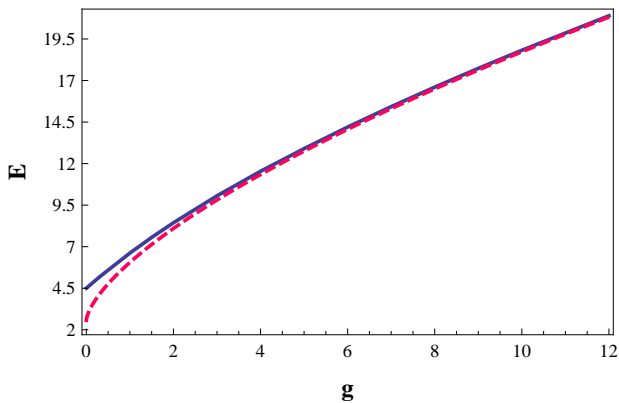}
\end{center}
\caption{\label{fffklolofghk3lg:beh} Comparison of the
approximations (\ref{app}) (broken curve) with the numerically exact
results (continuous curve).}
\end{figure}

\begin{figure}[h]
\begin{center}
\includegraphics[width=0.45\textwidth]{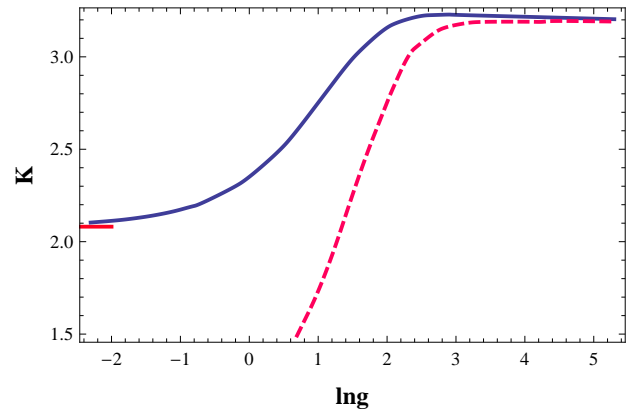}
\end{center}
\caption{\label{fffklolofgfffhk3lg:beh} Degree of correlations
calculated from the numerically exact  wavefunctions (continuous
curve) and the approximate ones (\ref{pps}) (broken curve), as a
function of $\ln g$. The horizonal line marks the
 result  determined as $g\rightarrow 0$  with
the use of
 (\ref{asym0}).}
\end{figure}

The system of Bose particles described by the Hamiltonian
(\ref{hamk}) gets fermionized for any $g\neq 0$ \cite{gi,astr} and
its ground-state wavefunction $\psi_{B}$ can be related to the
lowest energy antisymmetric
 wavefunction $\psi_{F}$ by
 \be
\psi_{B}(x_{1},x_{2},x_{3})=|\psi_{F}(x_{1},x_{2},x_{3})|.\label{opp}\ee
Therefore,  in the limit  $g\rightarrow 0$, $\psi_{B}$ tends to the
modulus of the Slater determinant \be \psi_{B}^{g\rightarrow
0}(x_{1},x_{2},x_{3})={1\over
\sqrt{3!}}|det_{n=0,j=1}^{2,3}(\varphi_{n}(x_{j}))|\label{asym0},\ee
where $\{\varphi_{n}\}$ are the single-particle orbitals of the
ideal system ($g=0$). In contrast, in the non-interacting case,
$\psi_{B}$ is nothing else but  the product function
$\psi_{B}^{g=0}=\prod_{i=1}^{3}\varphi_{0} (x_{i})$. The above in
turn implies  that the  quantities associated with $\psi_{B}$
exhibit generally a discontinuity at the point $g=0$. In particular,
when it comes to the  energy, it tends to $4.5$ as $g\rightarrow 0$,
while at $g=0$, it has the value $1.5$. In order to test the
applicability of  the approximations   (\ref{app}) and (\ref{pps}),
we determined numerically the ground-state bosonic wavefunction
$\psi_{B}$ and its corresponding energy, for a wide range of values
of $g$. The results of Eq. (\ref{app}) are compared with our
accurate numerical results in Fig. \ref{fffklolofghk3lg:beh}, from
which it can be seen that the approximate energy  tends from below
to the exact one with increasing $g$. Surprisingly, Eq. (\ref{app})
yields very good estimates of the true values already when $g$
exceeds the value $g=5$.
 However, the results of
Fig. \ref{fffklolofghk3lg:beh} are not a  good indicator of the
accuracy of  the approximate bosonic wavefunctions (\ref{pps}). To
gain insight into their range of applicability, we analyse their
ability to reproduce the numerically exact bechaviour of the degree
of correlation  \cite{degree} \be K=[tr\hat{\rho}^{2}]^{-1}=[\int
\rho(x,y)^2dx dy]^{-1}\label{lnearll}, \ee where $\rho$ is the RDM
expressed in coordinates \be \rho (x,y)=\int
\psi(x,x_{2},x_{3})\psi(y,x_{2},x_{3})dx_{2}dx_{3}.\ee
 The degree of correlation $K$ counts approximately
the number of
 orbitals actively involved in the Schmidt decomposition of the RDM and is one of the  transparent measures of
 correlation.
 It is worth stressing at this point that the linear correlation entropy $L$, which is also a popular measure of
correlation \cite{manz,lin3,monk,kosPLA,helium},  is related to $K$
via $L=1-1/K$.
 The results for the degree of correlation  calculated
 from the  numerically exact bosonic  wavefunctions  and the
approximate ones    $\psi_{B}^{ap}$ (\ref{pps}) are plotted in Fig.
\ref{fffklolofgfffhk3lg:beh}
  up to $g=200$.
  As one can see, acceptable results are reached   just at a value of
about $g= 200$ ($\ln200\approx5.29$).  To complete our discussion,
we compare  the densities $n(x)=3\rho(x,x)$ evaluated from the
approximate bosonic wavefunctions (\ref{pps}) with the exact ones
determined numerically.
\begin{figure}[h]
\begin{center}

\includegraphics[width=0.231\textwidth]{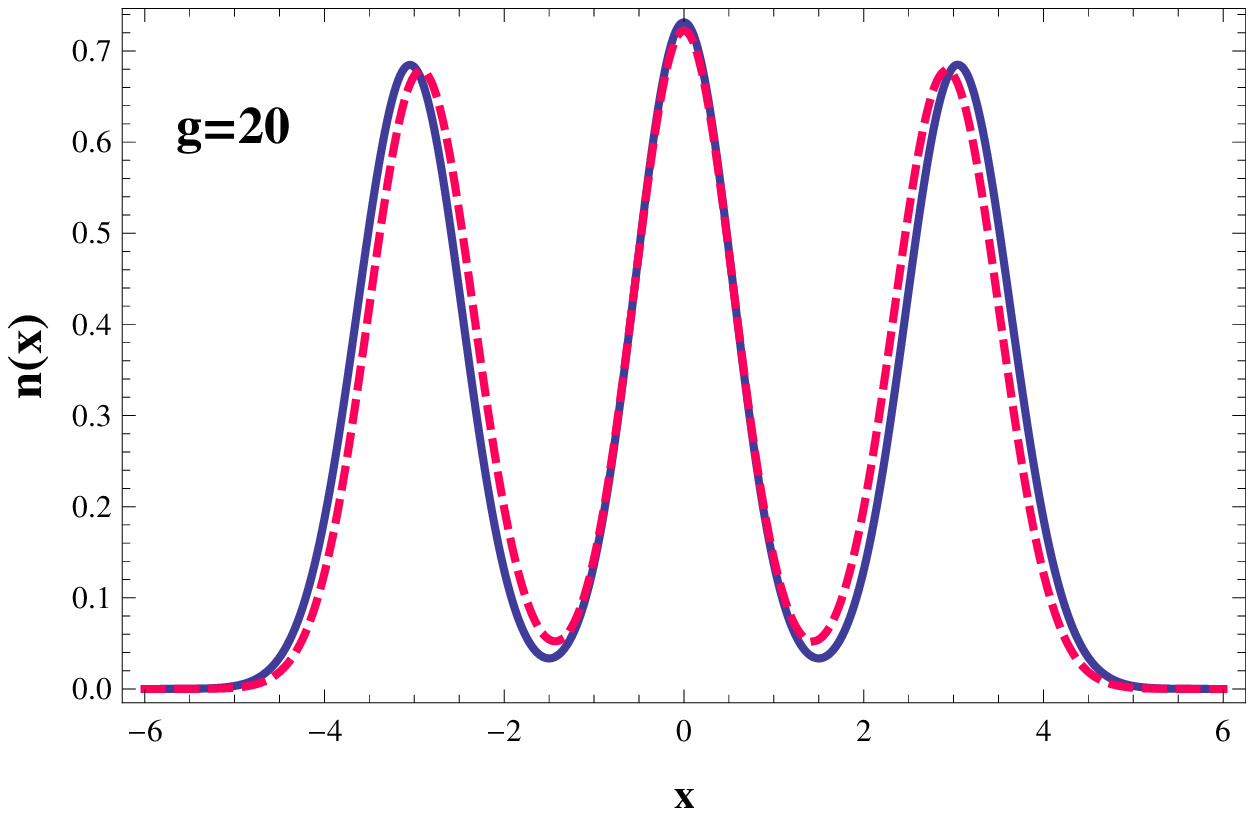}
\includegraphics[width=0.231\textwidth]{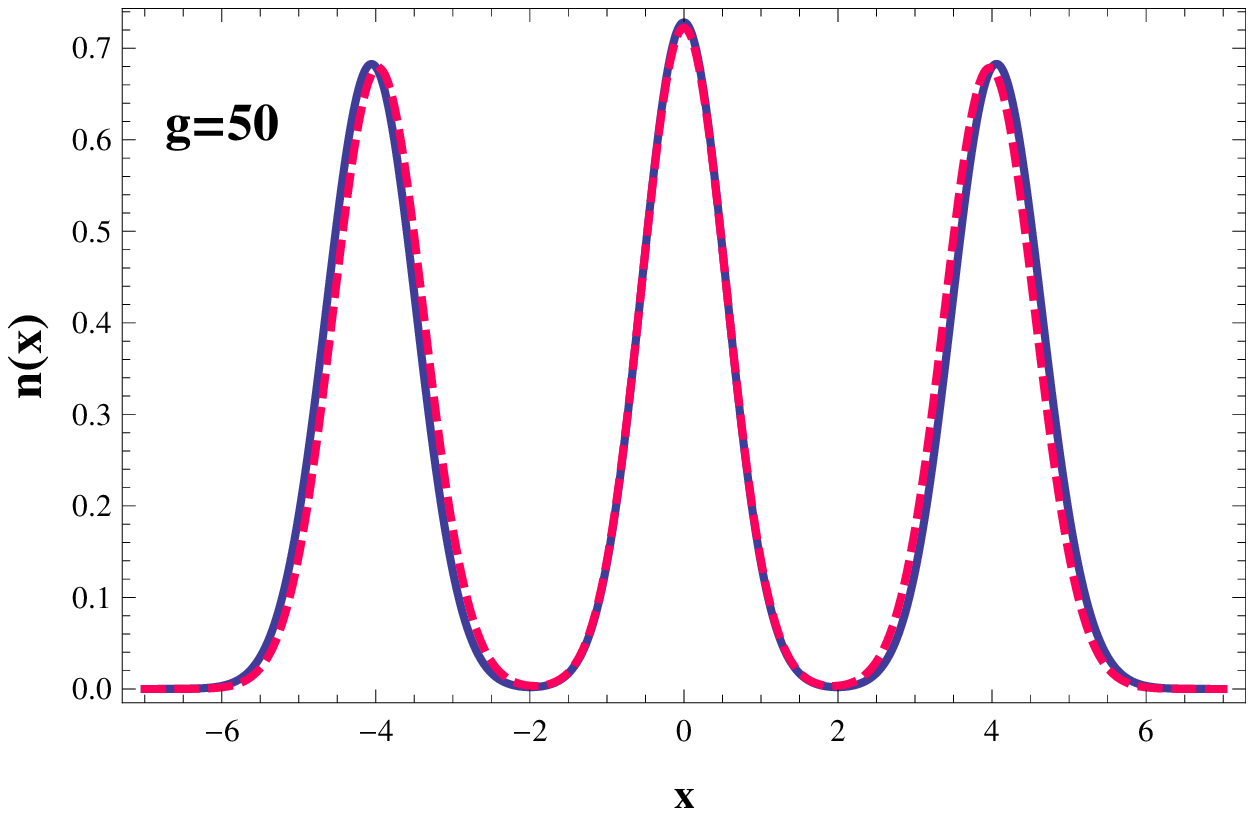}

\includegraphics[width=0.231\textwidth]{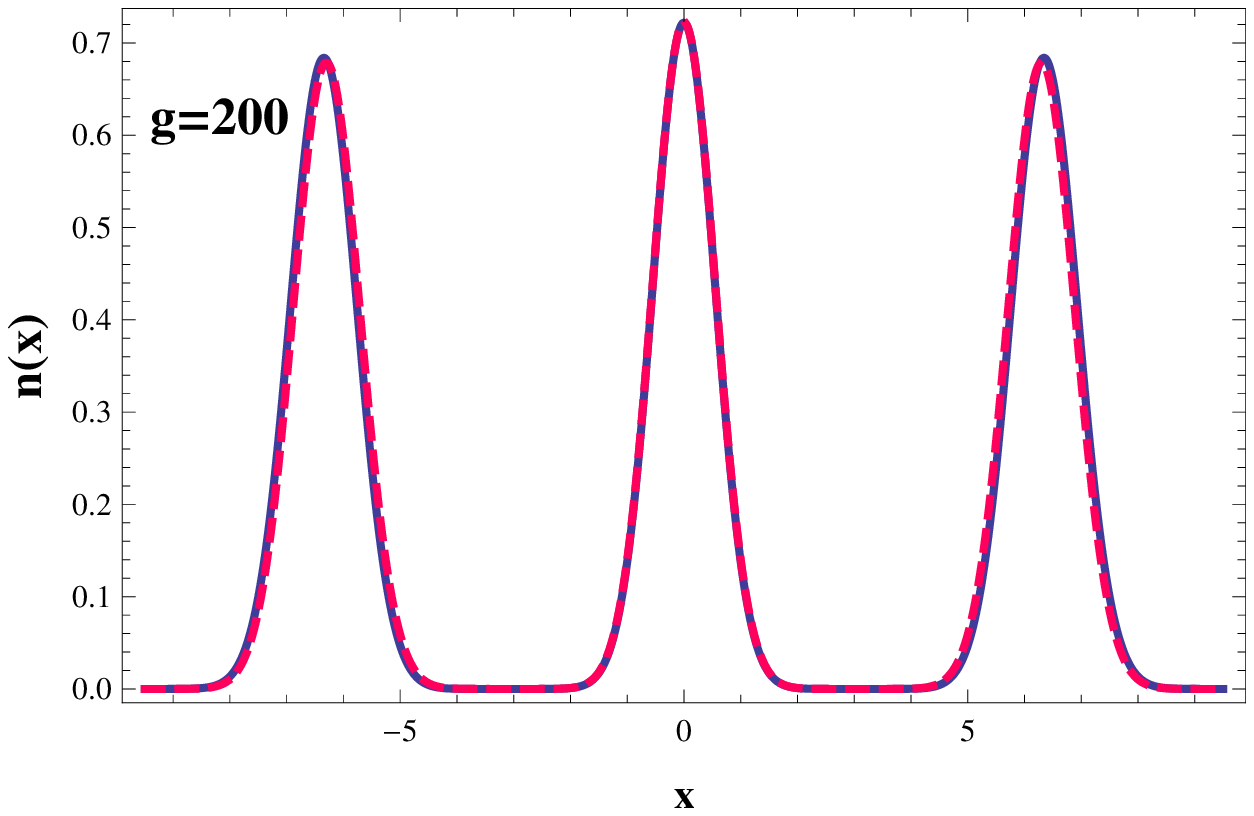}

\end{center}
\caption{\label{fdffklolofgrrhklg:beh} Comparison of
 one-particle density $n(x)=3\rho (x,x)$ calculated from the approximate bosonic wavefunction (\ref{pps}) (broken curve)
 with the
  exact
one determined numerically (continuous curve), for $g=20$
$(\ln20\approx 3)$, $g=50$ $(\ln50\approx 3.9)$, and $g=200$ $(\ln
200=5.29)$.}\end{figure}Our results are displayed in Fig.
\ref{fdffklolofgrrhklg:beh}, for three different values of the
interaction strength: $g=20,50$, and $g=200$.
  As one could have expected,
  the agreement between the approximate and exact densities
 is observed,  at least within the graphical accuracy, only in the last case.

 \section{Asymptotic expansion
of the RDM}\label{2} Now we come to the main goal of this paper,
which is to provide the Schmidt decomposition of the RDM in the
strong-correlation limit. To begin with our analysis, we take into
consideration the RDM for the
 approximate wavefunction $\psi_{B}^{ap}$ (\ref{pps}),
\begin{eqnarray}\rho_{ap}(x,y )=\int
\psi_{B}^{ap}(x,x_{2},x_{3})\psi_{B}^{ap}(y,x_{2},x_{3})
dx_{2}dx_{3}\label{pprr}\end{eqnarray} which becomes exact  as
$g\rightarrow \infty$ ($x_{c}\rightarrow \infty$). An easy
inspection of Eq. (\ref{pprr}) reveals that in this limit, it
reduces to the form \be\rho_{ap}^{g\rightarrow
\infty}=\rho^{g\rightarrow
\infty}=\rho_{1}+\rho_{2}+\rho_{3},\label{polo}\ee with  \be
\rho_{1}(x,y)=A\int\psi_{ap}(z_{1},x,z_{2})\psi_{ap}(z_{1},y,z_{2})dz_{1}dz_{2},\label{www}\ee
\begin{eqnarray}
\rho_{2}(x,y)=\nonumber\\=A\int\psi_{ap}(x+x_{c},z_{1},z_{2})\psi_{ap}(y+x_{c},z_{1},z_{2})dz_{1}dz_{2}\label
{41},\end{eqnarray} and \begin{eqnarray}
\rho_{3}(x,y)=\nonumber\\=A\int\psi_{ap}(z_{1},z_{2},x-x_{c})\psi_{ap}(z_{1},z_{2},y-x_{c})dz_{1}dz_{2}\label{51},\end{eqnarray}
where $A=2C_{\infty}^2$.  We have obtained exact closed-form
solutions for the above
 integrals. However, since they are quite lengthy,
 we report here only their numerical expressions
 \be \rho_{1}(x,y)\approx 0.2407
e^{-0.8944 x^2+0.1499 x y-0.8944 y^2}.\label{34}\ee
\begin{eqnarray} \rho_{2/3}(x,y)\approx \nonumber\\\approx 0.2262 e^{ -0.7618[(x \pm x_{c})^2+(y\pm x_{c})^2]+ 0.0770(x\pm x_{c})
(y \pm x_{c})},\label{jkld}\end{eqnarray} which does not limit the
generality of our further consideration.   It it worthwhile to note
here that the same asymptotic behaviour for
 the RDM  will be obtained for the case of   fermions.

Using Eq. (\ref{polo}),
  we can obtain a closed-form asymptotic expression for the
 one-particle density:
\begin{eqnarray} n^{g\rightarrow\infty}=3\rho^{g\rightarrow\infty}(x,x)\approx3(0.2407 e^{-1.6389
x^2}+\nonumber\\+0.2262e^{-1.4466(x-x_{c})^2}+0.2262e^{-1.4466(x+x_{c})^2}).\end{eqnarray}
An inspection of the computed $n^{g\rightarrow\infty}$ shows that it
exhibits exactly the Gaussian peaks   centred  at the classical
equilibrium positions of the particles.  It is worth emphasizing
that the peaks at $x=\pm x_{c}$ have the same
 profile.

 Being real and symmetric, the function ˜$\rho_{1}$
(\ref{34}) has the Schmidt decomposition \be \rho_{1}(x,y)
=\sum_{l=0}\lambda_{l}^{(1)}u_{l}(x)u_{l}(y), \label{pr}\ee where
$\{u_{l}\}$ and $\{\lambda_{l}^{(1)}\}$ are determined by \be
\int\rho_{1}(x,y) u_{l}(y)dy=\lambda_{l}^{(1)} u_{l}(x),\label{pepr}
\ee $\langle u_{l}|u_{k} \rangle =\delta_{lk}$. One can note that
the introduction of new coordinates in the functions $\rho_{2}$
(\ref{jkld}) and $\rho_{3}$ (\ref{jkld}) by \be x\mapsto
\tilde{x}-x_{c},y\mapsto \tilde{y}-x_{c},\label{fdf}\ee and \be
x\mapsto \tilde{x}+x_{c},y\mapsto \tilde{y}+x_{c},\label{fdf1}\ee
respectively, transforms them into $g$-independent forms which are
identical
 to each other,
 $\rho_{2}(x,y),\rho_{3}(x,y)\mapsto\tilde{\rho}(\tilde{x},\tilde{y})$,
 \begin{eqnarray}\tilde{ \rho}(\tilde{x},\tilde{y})\approx0.2262 e^{ -0.7618(\tilde{x}^2+\tilde{y}^2)+
 0.0770\tilde{x} \tilde{y} }.\label{jkld3}\end{eqnarray}
The above function is also real and symmetric, thus its Schmidt
decomposition is \be
\tilde{\rho}(\tilde{x},\tilde{y})=\sum_{l=0}\lambda_{l}^{(2)}v_{l}(\tilde{x})v_{l}(\tilde{y}),\label{fh}\ee
where $\{v_{l}\}$ and $\{\lambda_{l}^{(2)}\}$ are determined by \be
\int\tilde{\rho}(\tilde{x},\tilde{y})
v_{l}(\tilde{y})d\tilde{y}=\lambda_{l}^{(2)}
v_{l}(\tilde{x}),\label{fh1} \ee $\langle v_{l}|v_{k} \rangle
=\delta_{lk}$. By changing the variables back in (\ref{fh}) in
accordance with (\ref{fdf}), one gets the expansion of $\rho_{2}$ in
the form
 \be \rho_{2}(x,y)=\sum_{l=0}\lambda_{l}^{(2)}v_{l}(x+x_{c})v_{l}(y+x_{c}).\label{ll9}\ee
 On the other hand, the change of variables back in (\ref{fh}) in
accordance with (\ref{fdf1}), yields the expansion of $\rho_{3}$ as
 \be \rho_{3}(x,y)=\sum_{l=0}\lambda_{l}^{(2)}v_{l}(x-x_{c})v_{l}(y-x_{c}).\label{ll91}\ee
Obviously, the one-particle orbitals $v_{l}(x\pm x_{c})$
  satisfy $\langle
v_{l}(x\pm x_{c})|v_{k}(x\pm x_{c})\rangle=\delta_{lk}$. Finally,
substitution of Eqs. (\ref{pr}), (\ref{ll9}), and (\ref{ll91}) into
(\ref{polo}) gives, as  $g\rightarrow \infty$ ($x_{c}\rightarrow
\infty$),
\begin{eqnarray} \rho^{g\rightarrow
\infty}=\nonumber\\=\sum_{l=0}[\lambda_{l}^{(1)}u_{l}(x)u_{l}(y)+\lambda_{l}^{(2)}v_{l}(x+x_{c})v_{l}(y+x_{c})+\nonumber\\+
\lambda_{l}^{(2)}v_{l}(x-x_{c})v_{l}(y-x_{c})].\label{klj}\end{eqnarray}
  In  this limit, the family
$\{u_{l}(x),v_{l}(x-x_{c}),v_{l}(x+x_{c})\}$ forms a complete and
orthonormal set, since the integral overlaps $\langle
v_{l}(x+x_{c})|v_{k}(x-x_{c})\rangle$, $\langle
u_{l}(x)|v_{k}(x-x_{c})\rangle$ and $\langle
u_{l}(x)|v_{k}(x+x_{c})\rangle$ vanish for any $l,k$. We can
therefore
 recognize
  Eq. (\ref{klj}) as
the Schmidt decomposition of the asymptotic RDM $\rho^{g\rightarrow
\infty}$. Because the asymptotic natural orbitals
 $v_{l}(x+x_{c})$ and $v_{l}(x-x_{c})$ correspond
to the same occupancy $\lambda_{l}^{(2)}$, i.e.,  double
degeneracies in the spectrum of the RDM  $\rho^{g\rightarrow
\infty}$ occur, the Schmidt decomposition (\ref{klj}) fails to be
unique \cite{Ghirardi}. Before going further we stress that the
conservation of probability  for the asymptotic occupancies gives
$\sum_{l=0}[\lambda_{l}^{(1)}+2\lambda_{l}^{(2)}]=1$.
 For the sake
of completeness, we give below another form of the Schmidt expansion
of $\rho^{g\rightarrow \infty}$, different from that of Eq.
(\ref{klj}). To begin with, we extend the results of \cite{Ghirardi}
to the case of more than one point of double degeneracy. From the
orbitals $v_{l}(z+x_{c})$ and $v_{l}(z-x_{c})$, which correspond to
$\lambda_{l}^{(2)}$, we define the new orbitals to be \be
\eta_{l}(z)={v_{l}(z+x_{c})+v_{l}(z-x_{c})\over \sqrt{2}},\ee
 and
\be \tau_{l}(z)={v_{l}(z+x_{c})-v_{l}(z-x_{c})\over \sqrt{2}},\ee
that fulfill $\langle \eta_{l}|\tau_{l}\rangle=0$. In terms of them,
Eq. (\ref{klj}) can be rewritten as
\begin{eqnarray}
\rho^{g\rightarrow\infty}=\sum_{l=0}[\lambda_{l}^{(1)}u_{l}(x)u_{l}(y)+\lambda_{l}^{(2)}\eta_{l}(x)\eta_{l}(y)+\nonumber\\+\lambda_{l}^{(2)}\tau_{l}(x)\tau_{l}(y)],\label{ddd}\end{eqnarray}
Since in the limit as $g\rightarrow \infty$ ($x_{c}\rightarrow
\infty$) we have $\langle\eta_{l}|\eta_{k}\rangle=\delta_{lk}$,
$\langle\tau_{l}|\tau_{k}\rangle=\delta_{lk}$ and the integral
overlaps $\langle u_{l}|\eta_{k}\rangle$, $\langle
u_{l}|\tau_{k}\rangle$, $\langle \eta_{l}|\tau_{k}\rangle$ vanish
for any $l,k$, Eq. (\ref{ddd}) yields nothing other than a Schmidt
form different from Eq. (\ref{klj}).

 The integral equations (\ref{pepr}) and
(\ref{fh1}) can easily be solved through a discretization technique
(see for example \cite{kosPLA}). The two lowest asymptotic
occupancies  are found numerically to be
$\lambda_{0}^{(1)}\approx0.3193, \lambda_{0}^{(2)}\approx0.3249 $.
Because the sum of all the remaining asymptotic occupancies,\\
$\sum_{l=1}[\lambda_{l}^{(1)}+2\lambda_{l}^{(2)}]$, is only about
$0.03$, it follows that in (\ref{klj}) and in (\ref{ddd}) only the
terms with $l=0$ are important and,  in consequence,
$\rho^{g\rightarrow\infty}$ approaches the form
\begin{eqnarray} \rho^{g\rightarrow
\infty}\approx\nonumber\\\approx\lambda_{0}^{(1)}u_{0}(x)u_{0}(y)+\lambda_{0}^{(2)}v_{0}(x+x_{c})v_{0}(y+x_{c})+\nonumber\\
+\lambda_{0}^{(2)}v_{0}(x-x_{c})v_{0}(y-x_{c}),\label{klffffj}\end{eqnarray}
in particular.

\begin{figure}[h]
\begin{center}
\includegraphics[width=0.45\textwidth]{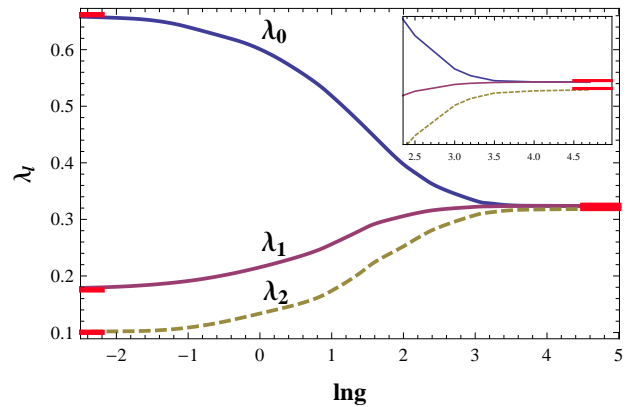}
\end{center}
\caption{\label{fff9klolofgff44fhk3lg:beh}  The dependence of the
three lowest  occupancies on $\ln g$. The inset highlights their
changes for large values of $g$. The asymptotic values are marked by
horizontal lines. The values of asymptotic occupancies
$\lambda_{0}^{(2)}=\lambda_{0}^{g\rightarrow\infty}=\lambda_{1}^{g\rightarrow\infty},\lambda_{0}^{(1)}=\lambda_{2}^{g\rightarrow\infty}$
are reported in the text, whereas, the values of three lowest
occupancies determined  as $g\rightarrow 0$ with the use of
(\ref{asym0}) are found numerically to be
$\lambda_{0}^{g\rightarrow0}\approx0.6619$,
$\lambda_{1}^{g\rightarrow0}\approx0.1755$,
$\lambda_{2}^{g\rightarrow0}\approx0.1004$.  }
\end{figure}

We close our discussion with Fig. \ref{fff9klolofgff44fhk3lg:beh},
which shows the numerically determined behaviour of $\lambda_{l}$,
for $l=0-2$ as a function of $\ln g$. It is seen how the occupancies
converge to their asymptotic values determined
 by the integral equations (\ref{pepr}) and
(\ref{fh1}), which confirms their validity. In particular, we
observe   how $\lambda_{0}$ and $\lambda_{1}$ converge  to an
asymptotic doublet.

\section{Summary}\label{3} In conclusion, we investigated the
ground-state correlation properties of the system composed of  three
charged bosons in a 1D
  harmonic trap.  Using the harmonic approximation
we explained the nature of the degeneracies appearing as
$g\rightarrow \infty$  in the spectrum of the
 RDM. An explicit representation of the
asymptotic natural orbitals and their occupancies has been derived
in terms of some $g$-independent integral equations.   Among other
results, we found that in the $g\rightarrow\infty$ limit the
occupancies $
\lambda_{0}^{(2)}=\lambda_{0}^{g\rightarrow\infty}=\lambda_{1}^{g\rightarrow\infty}\approx0.3249$,
$\lambda_{0}^{(1)}=\lambda_{2}^{g\rightarrow\infty}\approx0.3193$
are the  only three that have  considerable values.  In other words,
it turned out that only the  three natural orbitals contribute
significantly to the asymptotic bosonic ground-state. We also
determined numerically the three lowest occupancies as functions of
$g$ and showed how they tend to their asymptotic values. In
particular, we obtained  a closed-form asymptotic expression for the
one-particle density given as a linear combination of Gaussian
functions centred at the classical equilibrium positions of the
particles.

 It would be interesting to fully investigate  the
effect of the number of particles on the correlation properties in
1D systems of strongly interacting bosons and/or spin fermions with
a Coulomb interaction. To the best of our knowledge, there is still
a lack of studies along this line. We hope our results will
stimulate others to undertake the investigation of this issue. This
will be also a subject of our further research.

\bibliography{aipsamp}

\end{document}